\documentclass[twocolumn,preprintnumbers,amsmath,amssymb]
{revtex4-2}
\usepackage{graphicx}
\usepackage[normalem]{ulem}
\usepackage[svgnames]{xcolor}
\usepackage{bm}
\usepackage{multirow}\usepackage{float}
\usepackage{titlesec}
\usepackage{xcolor}

\begin{document}
\title{Thermal diffusivity and its lower bound in orthorhombic SnSe}

\author{Valentina Martelli$^{1}$}
\author{Fabio Abud$^{2}$}
\author{Julio Larrea Jiménez$^{1}$}
\author{Elisa Baggio-Saitovich$^{3}$}
\thanks{}
\author{Li-Dong Zhao$^{4}$}
\author{Kamran Behnia$^{5}$}

\affiliation{(1) Laboratory for Quantum Matter under Extreme Conditions,  Instituto de Física, Universidade de São Paulo, 05508-090, São Paulo, Brazil\\
(2) Instituto de Física, Universidade de São Paulo, 05508-090, São Paulo, Brazil\\
(3) Centro Brasileiro de Pesquisas Física (CBPF)\\ 
(4) School of Materials Science and Engineering, Beihang University, Beijing 100191, China\\
(5) Laboratoire de Physique et d'Étude des Matériaux\\ (ESPCI Paris - CNRS - Sorbonne Universit\'e), PSL Research University, 75005 Paris, France\\}

\date{\today} 
\begin{abstract}
The orthorhombic monochalcogenide SnSe has attracted much attention in recent years as a promising high-temperature thermoelectric material. We present a study of  its thermal conductivity and specific heat of SnSe between  2~K and 300~K  and quantify its anisotropic thermal diffusivity, $D$. For both crystallographic orientations, thermal diffusivity remains above the recently identified  Planckian limit ($D >  v_s^2 \tau_P$, where $v_s$ is the sound velocity and $\tau_P= \hbar/k_BT$) and its anisotropy in $D$ is set by the anisotropy of $v_s$. Comparison with cubic members of the IV-VI family leads to a consistent picture, where the diffusivity in all members of the family is set by the product of v$_s$, $\tau_P$ and the 'melting' velocity derived from the melting temperature.
\end{abstract}
\maketitle

\section{INTRODUCTION}

Like elemental black phosphorus, SnSe crystallizes in an orthorhombic  crystal structure. A member of group-IV monochalcogenides, it is a promising material with potential applications in fields ranging from solar cells and  batteries to supercapacitors~\cite{Kumar2021}. Its properties the monolayer limit are also attracting newt attention~\cite{barraza2021colloquium,huang2017liquid_exfol_SnSe, gomes2015phosphorene_analogue,wang2017_2D_multiferroic, wu2016_2D_SnSe_ferroics,fei2015giant_piezo_monoIV,titova2020_reviewMonochalcogenides}. The bulk solid became a well-known thermoelectric material  following the report of a large figure of merit driven by its low thermal conductivity~\cite{Zhao2014ultralow,chang2018_3D-charge_and2D-phonon_SnSe, chang2018thermoelectric_surpriseSnSe}. The magnitude of thermal conductivity became a subject of controversy, with  a large variety in its reported amplitudes
\cite{wei2016_intrinsic,ibrahim2017, wang2018_pdopedSnSe, sassi2014Thermoel_ptypeSnSe,serrano2015_poly-nanostr-SnSe,li2015,heremans2015anharmonicity}. 

SnSe is a layered solid with easy-cleaving $b-c$  planes in which atomic bonds are stronger than in the perpendicular orientation generating a significant anisotropy in structural properties. This orthorhombic crystal structure, found in other binary IV-VI salts (like GeSe) can be viewed as a distorted rock-salt cubic structure of PbTe, another member of this family \cite{Littlewood1980}.  The competition between the rock-salt structure and the two (orthorhombic and rhombohedral) less symmetric options in this family is driven by an interplay between Peierls distortion, s-p hybridization and spin-orbit coupling and has been a longstanding subject of meditation, contemplation and debate~\cite{Littlewood1980,Burdett1984,Seo1999,Peierls_book,behnia2016finding_merit}. 

Planckian time ($\tau_{p}=\hbar/k_B T$, with $k_B$ Boltzmann constant and $T$ the absolute temperature) as a boundary to rate of dissipation has attracted much attention in recent years. In 2013, Bruin \textit{et al.} ~\cite{bruin2013} noticed that  the effective scattering time of electrons in a variety of metals when electrical resistivity is linear in temperature is of the order of $\tau_P$. The behavior of phonons in insulators came under scrutiny a few years afterwards and it was found that when phonon-phonon scattering becomes dominant and \textit{thermal} resistivity is $T$-linear, the phonon scattering time  approaches $\tau_P$ without falling below it~\cite{martelli2018, behnia2019_lowerbound}. This experimental observation raised a  question. Above the Debye temperature, phonons obey the Boltzmann-Maxwell distribution. Therefore, how can the Planck constant infer with their dissipation rate?  Mousatov and Hartnoll~\cite{mousatov2020_planckianbound} argued that this lower bound to thermal diffusivity is set by the energy scale of the  melting temperature of the crystal. They proposed a universal scaling relation between the ratio of the phonon scattering time to the Planckian time and the ratio of the 'melting' velocity to the sound velocity. 

The previous experimental observations~\cite{martelli2018, behnia2019_lowerbound,xu2021_In2O3} and the theoretical scenarios~\cite{mousatov2020_planckianbound,Tulipman2020} were focused on cubic insulators and ignored the issue of anisotropy. In this paper, we will present and examine the case of orthorhombic, and therefore anisotropic, SnSe. By quantifying the thermal diffusivity  and sound velocity along the $b$ and $c$ directions, we will examine the account of the amplitude and anisotropy of thermal diffusivity in  the Moussatov-Hartnoll picture.

\section{EXPERIMENTAL} 

High-quality SnSe crystals were synthesized using a modified vertical Bridgman method \cite{chang2018_3D-charge_and2D-phonon_SnSe}. High-purity elemental constitutes were measured and loaded into carbon-coated conical silicon tubes and the crystals were grown in a temperature gradient from 1213~K to 973~K at a slow rate of 1~K/h. Finally, SnSe crystals with diameter of 12~mm and length of 30~mm were obtained. 
The investigated single crystals display an $n$-type behaviour, with a carriers density in the order of $10^{16}$~cm$^{-3}$.
The contribution of charge carriers to thermal transport can be estimated by Wiedemann-Franz law and remains negligible in our range of measurements.

\begin{figure}
\centering
\includegraphics[width=0.45\textwidth]{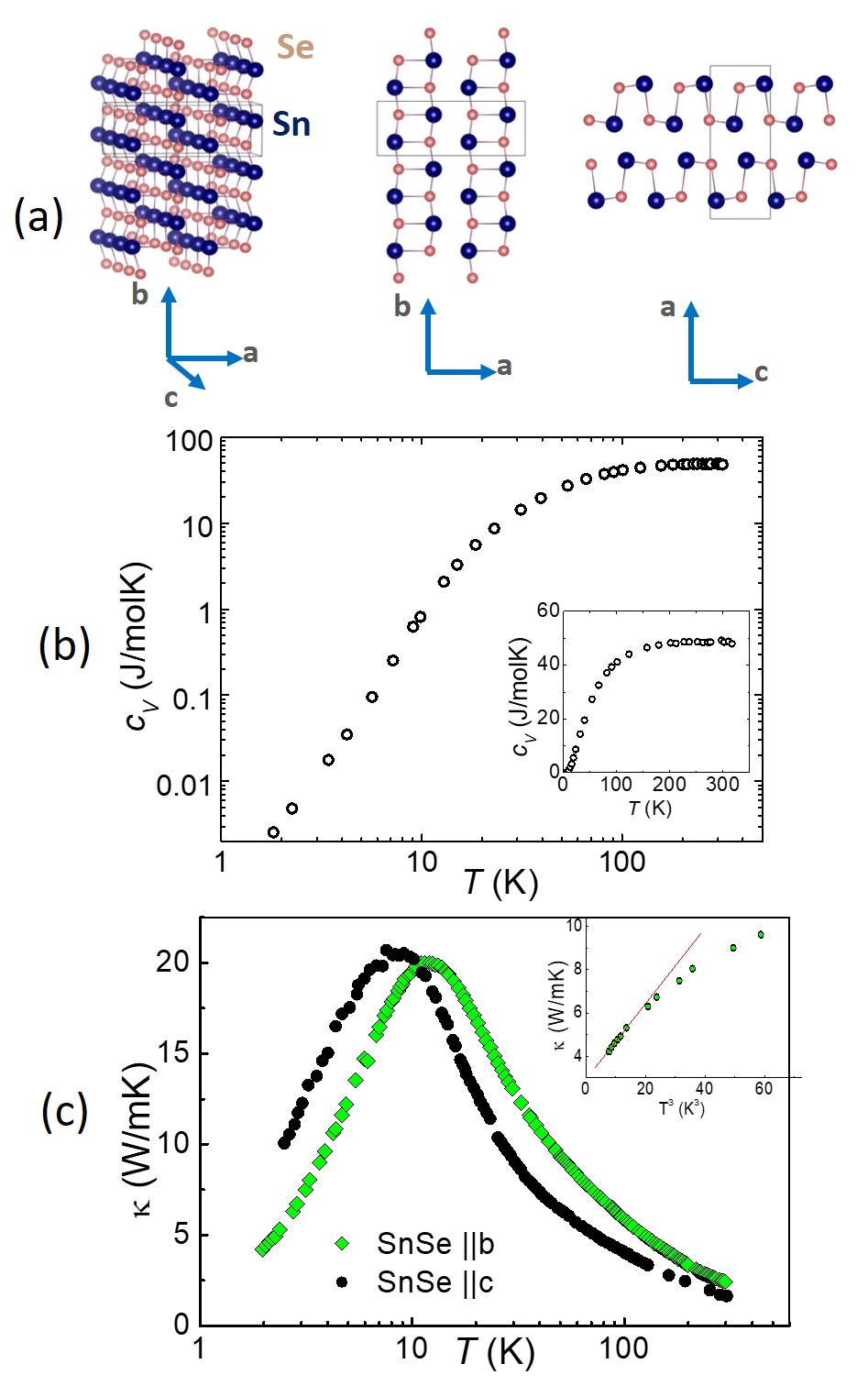}
\caption{\textbf{Thermal conductivity and specific heat of SnSe:} (a) The projections of the main crystallographic directions of the orthorhombic SnSe. Lattice parameters from Ref. \cite{chattopadhyay1986}.  (b) Specific heat (2-300~K) of SnSe. The inset shows the same data set in a linear scale. (c) Thermal conductivity of SnSe as a function of temperature between 2-300K when a thermal gradient is applied along the $b$ and $c$ directions.}
\label{fig:Fig1}
\end{figure}

The specimens were aligned and cut appropriately to apply a thermal gradient along the in-plane $b$ and $c$ crystallographic directions, obtaining a geometric factor $g=\frac{S}{L}=1.5$~mm where $S$ is the cross section area and $L$ is the separation between the leads that probe the temperature gradient. We measured thermal conductivity with a standard two-thermometers one-heater setup in the 2-300~K temperature range. Heat capacity was also determined between 2 and 300~K, using a standard platform in a Quantum Design cryostat. The molar heat capacity is divided by the mass of the specimen and multiplied by the molar mass of SnSe, $M_m=197.67$~g/mol.  The density of the single crystals was determined by combining volume and mass measurements $\rho=(6.4\pm 0.4)$~g.cm$^{-3}$ that inside the experimental uncertainty is in agreement with the expected value for a fully dense sample. 

\begin{figure}
\centering
\includegraphics[width=0.45\textwidth]{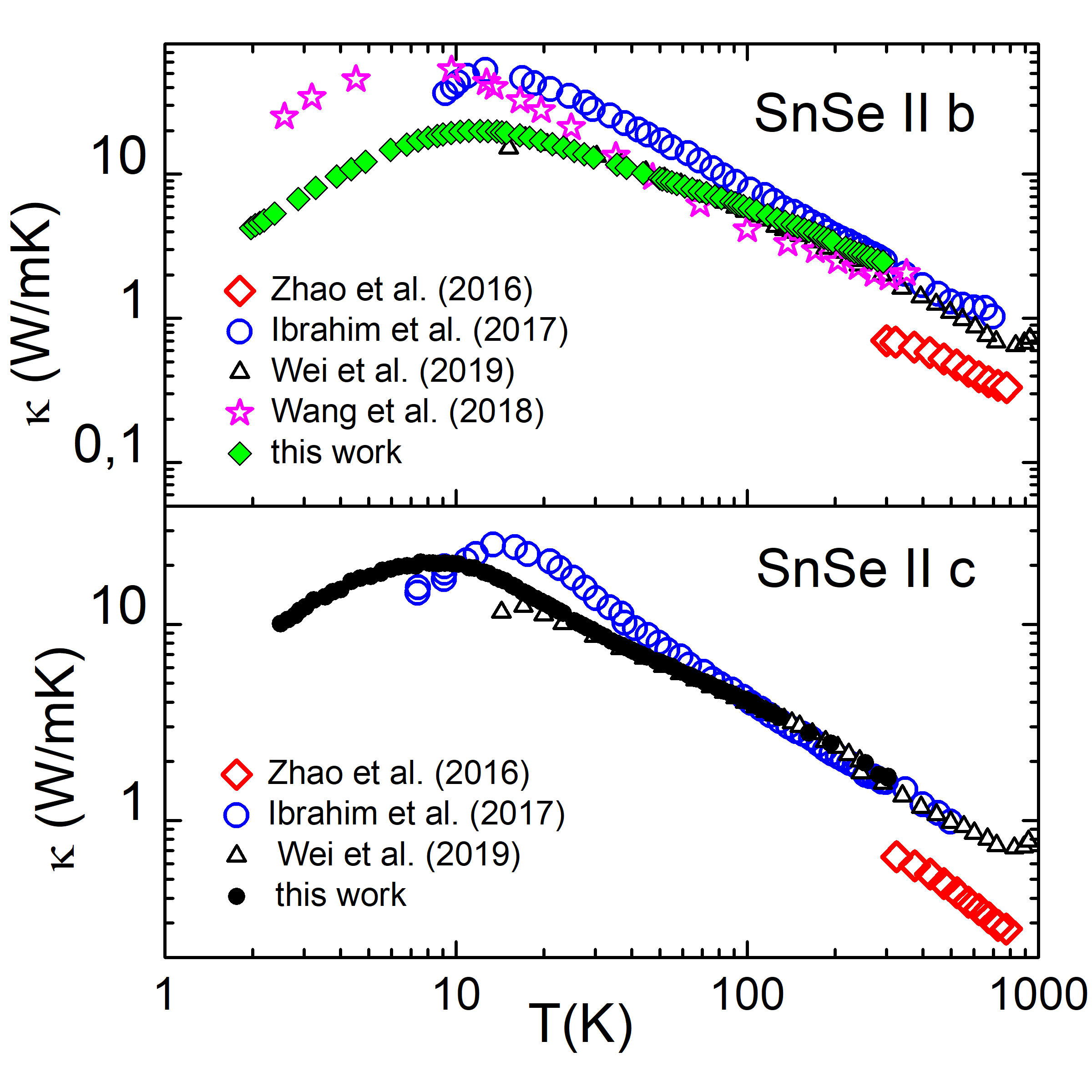}
\caption{\textbf{Thermal conductivity of SnSe}: Extended temperature range thermal conductivity of SnSe with heat conduction along the $b$ and $c$ direction, in the upper and lower panel respectively. Comparison between this work and previously reported data \cite{Zhao2014ultralow, ibrahim2017,wang2018_pdopedSnSe, wei2019thermoelectric_fullydense_singleSnSe}.}
\label{fig:Fig2}
\end{figure}

\section{RESULTS}

Fig \ref{fig:Fig1}(b) shows the temperature dependence of the  specific heat data $C_v(T)$, to be compared with what was reported previously~\cite{sassi2014Thermoel_ptypeSnSe,Zhao2014ultralow,ibrahim2017}.  We find that 
above $\sim$ 200~K the specific heat plateaus to  a value of about $\sim$ 50 J.K$^{-1}$.mol$^{-1}$, consistent with the expected Dulong-Petit value of 50 J.K$^{-1}$.mol$^{-1}$ for a solid with bi-atomic crystal cell. 
The Debye temperature $\theta_D$, determined by fitting the  specific heat contribution $\beta$ in $C_v/T=\gamma  + \beta T^2$ and quantifying $\beta$, was found to be  $\theta_D\approx 190K$, consistent with previous experimental data \cite{sassi2014Thermoel_ptypeSnSe}. The theoretically expected  value has been calculated to be $\theta_D\approx 140~K$  \cite{carrete2014}. 




As seen in Fig.\ref{fig:Fig1}(c), for both directions thermal conductivity displays the typical behaviour expected for an insulator. Thermal conductivity, $\kappa$, peaks at intermediate temperatures for both orientation. The peak height is about 20 W/m.K  for both directions. The peak occurs at 12K for $b$ direction and at 9K for the $c$ direction. At low temperatures, below the peak, $\kappa$ displays a $T^3$ power law (see inset). 

As seen in  Fig. \ref{fig:Fig2}, according to our data, at room temperature, the amplitude of $\kappa$ is  2.4~W/m.K and 1.6~W/m.K along the $b$ and $c$ axes, respectively.  A comparison of our data with previous publications indicates that our result is in good agreement with most works on single crystals~\cite{ibrahim2017,wang2018_pdopedSnSe,wei2019thermoelectric_fullydense_singleSnSe}, but  differs by a factor ~3 with respect to the reported ultra-low values~\cite{Zhao2014ultralow}. The discrepancy among different data sets was debated previously~\cite{wei2016_intrinsic}. Our results on fully dense single crystalline samples reaffirm that likely the  thermal conductivity of single crystalline SnSe are not as low as initially claimed and the discrepancy is a matter of accurate measurement.

Our experimental results can also be compared with \textit{Ab-initio} calculations in both the low temperature orthorhombic phase ($Pnma$, $T<800~T$) \cite{carrete2014} and in the high-temperature base-centered orthorhombic phase ($Cmcm$, $T>800~T$) \cite{aseginolaza2019phonon_collapse_SnSe}. Since the structural phase transition is of second order, one expects a smooth evolution of the physical properties across it. Both theoretical works \cite{carrete2014,aseginolaza2019phonon_collapse_SnSe} are in agreement with the results presented in this work and in other previous experimental studies \cite{ibrahim2017, wei2019thermoelectric_fullydense_singleSnSe}. This settles the magnitude of the intrinsic thermal conductivity in single-crystalline SnSe to be $ \sim$ 2 W/m.K at room temperature.

The amplitude of the peak in $\kappa$ around 10~K varies between 20 and 60 W/m.K, indicative of difference in the maximum phonon-mean-free-path across samples. At high temperature, thermal conductivity  follows a  $1/T$ behavior with a clear anisotropy between the two crystallographic orientations, in agreement with what previously reported~\cite{ibrahim2017, wei2019thermoelectric_fullydense_singleSnSe}. 

\begin{figure}
\centering
\includegraphics[width=0.45\textwidth]{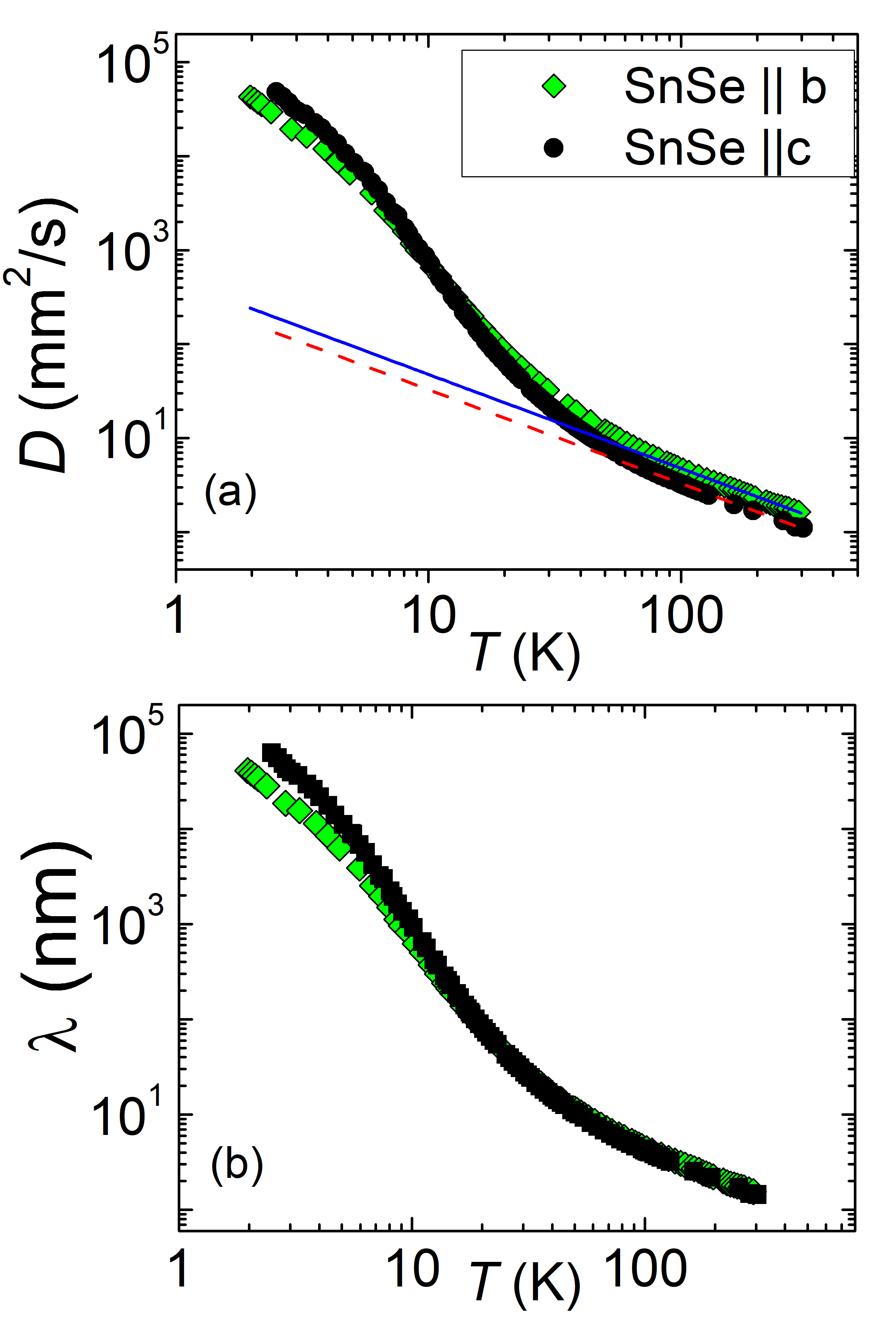}
\caption{\textbf{Anisotropic diffusivity:} (a) Thermal diffusivity and (b) mean free path of SnSe for $b$ and $c$ crystallographic directions. The solid and dashed lines in (a) represent $T^{-1}$ power law.}
\label{fig:Fig3}
\end{figure}
Thermal diffusivity $D$ of a solid is defined by the heat equation
$\frac{\partial T}{\partial t}=D \frac{\partial^2 T}{\partial x^2}$, where thermal transport is supposed to be established along a certain direction (for instance $x$). Combining the Fourier equation with the conservation of energy, one finds that $D$ is  the ratio of thermal conductivity and specific heat per volume. Fig. \ref{fig:Fig3} (a) shows the obtained $D$  along $b$ and $c$ orientations.  One can see that they cross around 10~K. Above 100~K, they display  a $\sim T^{-1}$ behaviour.  The phonon mean free path ($\lambda$) along the two orientations is shown in fig. \ref{fig:Fig3}(b), extracted using $\kappa=\frac{1}{3} c_{ph} v_s \lambda$. The sound velocity $v_s$  for each direction was determined as discussed in the Appendix.  At the lowest investigated temperature, $\lambda \sim 0.05$~mm, still more than one order of magnitude lower than the smallest geometrical dimension of the sample ($\sim$2 mm), indicating that phonons do not reach the ballistic regime.  



\begin{figure}
\centering
\includegraphics[width=0.45\textwidth]{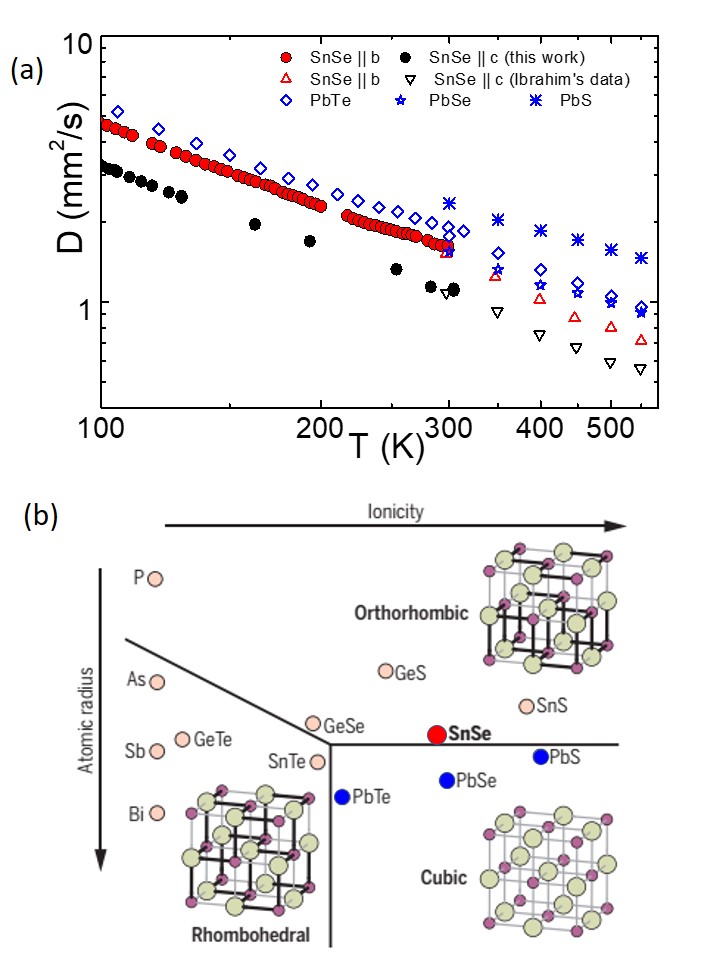}
\caption{\textbf{High-temperature diffusivity:} (a) High-temperature diffusivity of SnSe compared with other calchogenides. Values for PbTe, PSe and PbS are obtained from \cite{parkinson1954molar, pashinkin2009heat,el1983thermophysical, morelli2008intrinsically}; SnSe \cite{ibrahim2017}. (b) Structural phase diagram of IV-VI salts (adapted from Ref. \cite{Littlewood1980,behnia2016finding_merit}).}
\label{fig:Fig4}
\end{figure}

Fig. \ref{fig:Fig4} compares our thermal diffusivity with previous studies of single crystalline SnSe and with three cubic members of IV-VI family. We can see that our results agree with the data reported by Ibrahim \textit{et al.}~\cite{ibrahim2017}.

\section{DISCUSSION}
 
It is instructive to compare the amplitude of the maximum  $\kappa$ in SnSe and in black phosphorus, which are iso-structural. The 20-60 W/m.K peak in SnSe is more than  one order of magnitude smaller than what was observed black phosphorus~\cite{machida2018}. In the latter system, below the peak temperature, the  phonon mean-free-path is comparable to the thickness  and the amplitude of the thermal conductivity is size-dependent~\cite{machida2018}. In contrast, in SnSe, in the ballistic regime is not attained. This implies the presence of disorder spatially extended enough to scatter long wavelength phonons. The absence of ballistic phonon transport hinders the possible emergence of phonon hydrodynamics~\cite{beck1974} driven by abundance of momentum-conserving phonon-phonon collisions~\cite{martelli2018,machida2018,machida2020}.  

\begin{figure*}
\centering
\includegraphics[width=0.85\linewidth]{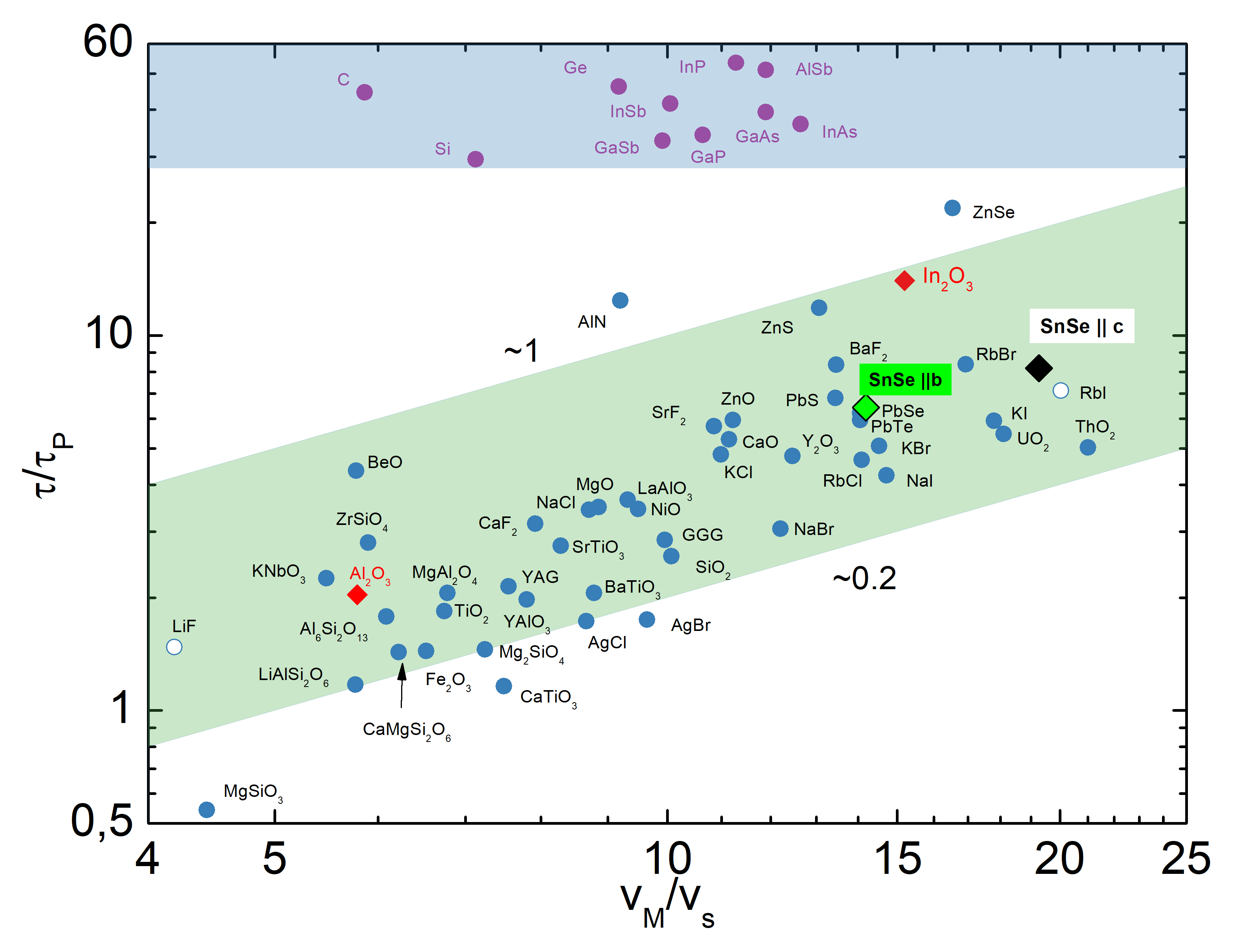}
\caption{\textbf{Scattering time vs sound velocity:} $\tau/\tau_p$ vs $v_M/v_s$ for different materials (Figure adapted from Ref. \cite{xu2021_In2O3}) compared with SnSe data of this work. The new points of SnSe are determined taking into account, as the $v_s$, the longitudinal sound velocity for each direction.}
\label{fig:Fig5}
\end{figure*}
We now turn our attention to the thermal diffusivity. The existence of a lower bound to thermal diffusivity $D$ in insulators~\cite{martelli2018, behnia2019_lowerbound} was noticed through the scrutiny of the slope of $D$ as a function of inverse of temperature in various solids. The  phonon scattering time  $\tau$ can be extracted, using:
\begin{equation}
    D= v_s^2 \tau
    \label{eq:D}
\end{equation}
It was found that even in the least conducting known materials $\tau$  approaches, but does not fall below the Planckian time $\tau_{p}$:
\begin{equation}
    \tau= s \tau_{p} = s \left[ \frac{\hbar}{k_B T } \right]
    \label{eq:s}
\end{equation}
where $s$ is a material-dependent constant that was found to be larger than unity in all known materials.

Mousatov and Hartnoll \cite{mousatov2020_planckianbound} proposed that the proximity to the Planckian limit is controlled by the sound velocity $v_s$ constrained by the bound $v_s\leq v_M$, where $v_M$ is the crystal melting velocity. The latter is defined as $v_M\equiv \frac{k_B T_M}{\hbar} a$ where $T_M$ is the melting temperature and $a$ is the lattice spacing.  The closer is $v_s$ to $v_M$, the closer is a certain the material to the lower bound. The melting velocity defined by the melting temperature and the atomic spacing sets an upper boundary to the velocity in a crystal. They showed that  for a large number of compounds $\tau/\tau_p$ and $v_M/v_s$ display a roughly linear correlation~\cite{mousatov2020_planckianbound}. More recently, Xu \textit{et al.}~\cite{xu2021_In2O3} measured the thermal conductivity and diffusivity of  cubic In$_2$O$_3$ and found that the properties of this compound fit in this picture.

The case of  orthorhombic SnSe provides an opportunity to probe the role of  the anisotropy of the sound velocity in this picture.  Using the components of the elastic tensor  experimentally reported in ref.~\cite{karunarathne2018}, we calculated the longitudinal sound velocities along $b$ and $c$ to be, respectively $v_{l,b}=$3146~m/s and $v_{l,c}=$2317~m/s (see Appendix).  Remarkably, the ratio  of the velocities along the two orientation corresponds to the anisotropy of thermal diffusivity in the intrinsic regime. Both thermal diffusivity and sound velocity are roughly 1.3 times larger along the b-axis compared to the c-axis. 

We note that the amplitude of sound velocity is compatible with the moderate atomic mass of atoms in SnSe. According to a recent expression for the speed of sound $v_s=\frac{\alpha}{c}\sqrt{(m_e/2M)}$~\cite{trachenko2020_speedofsound}, where $c$ is the light velocity, $\alpha$ is the fine structure constant and $m_e$ and $M$ are the masses of bare electron and the atom, respectively. Injecting the mass of Sn (119 atomic mass units), one finds 3500 m/s.

Taking the high temperature slope of the diffusivity $D\cdot k_B/(\hbar v_s^2)$ as a function of $T^{-1}$ (see straight lines in Fig. \ref{fig:Fig3} (a)) one can quantify the parameter $s$ for both directions considering the computed value for the direction-dependent sound velocity $v_s$.  We found that $s_{||b}=6.3$ and $s_{||c}=8$. We did not measure thermal conductivity along the a-axis, but using the data reported by Ibrahim et al., one obtains  $s_{||a}=4$.

The phonon lifetime according to Raman spectroscopy is $\tau\sim$ 0.1 ps at 800K\cite{liu2018phonon_SnSe_Raman}; using this time, one finds $\tau/\tau_P\sim 10$ in good agreement with what is found here.

The melting velocity $v_M$ of SnSe can be calculated from the melting temperature $T_M=1134~K$ \cite{okamoto1998_SnSe} and the average interatomic distance $a=\sqrt[3]{\frac{M_m}{N_A n_m}}$ where $M_m$ is the molar mass (197.67~g/mol), $N_A$ is the Avogadro number and $n_m$ = 2 is the number of atoms per unit formula in SnSe. We obtain $v_M=44658$~m/s for SnSe. This allows us to put SnSe on Fig. \ref{fig:Fig5} (re-adapted from Ref. \cite{xu2021_In2O3}) together with a large number of systems previously tabulated \cite{mousatov2020_planckianbound}. One can see that along b-axis, SnSe is close to the cubic members of the IV-VI family. The larger $s$ for the c-axis corresponds can be explained by a lower sound velocity along this direction, which amplifies both  $\tau/\tau_p$ and $v_M/v_s$. 

Let us note that $\frac{D}{v_s^2 \tau_P}\propto \frac{v_M}{v_s}$  can be rewritten as $\frac{D}{v_s}\propto \tau_P v_M$. The right hand side is isotropic implying that the anisotropy of $D$ and $v_s$ should cancel out. As one can see in figure \ref{fig:Fig5}, for the 4 members of the IV-VI family examined here, the proportionality factor is close to 0.5. As a consequence, the expression $D\approx 0.5 v_sv_M\tau_P$ gives a satisfactory account of the magnitude of $D$ for both orientations in SnSe as well as the cubic members of this family, confirming the relevance of the picture put forward in ref.~\cite{mousatov2020_planckianbound} to members of this family.

The anisotropy of thermal diffusivity appears to be driven by the anisotropy of sound velocity.  The intrinsic phonon anharmonicity due to the proximity of a lowering-symmetry structural transition \cite{li2015,Littlewood1980,behnia2016finding_merit} plays a major role. Moreover, sound velocity depends on the elastic constants, whose link to phonon anharmonicity was discussed previously \cite{karunarathne2021_negativeExpansionSnSe}. Recently, Wu  and Sau~\cite{Wu2021} have proposed a heat transport model in a strongly anharmonic system
in which the thermal diffusivity can become lower than the Planckian bound. Our results indicate that SnSe in spite of its low thermal diffussivity and strong anharmoinicity respects the Planckian bound.

We end our discussion by comparing the magnitude of thermal diffusivity in this solid with the Universal lower bounds to thermal diffusivity of liquids~\cite{Trachenko2021}. If $D$ decreases linearly down to the melting temperature of 1138 k, it will become as small as $\approx$0.2~mm$^2$/s  at the onset of melting. This is still one order of magnitude larger than the lower boundary ($D_{min}=\frac{1}{4\pi}\frac{\hbar}{m_e M}$)~\cite{Trachenko2021} in a supercritical fluid with an atomic mass of $M\approx$ 100 amu.

In summary, we studied the anisotropic thermal diffusivity of SnSe for two crystallographic orientations between the 2~K and 300~K. We found that our experimental findings matches a  recent theoretical framework provided that one takes into account the anisotropy of sound velocity.


VM acknowledges the support of JP-FAPESP (2018/19420-3). JLJ acknowledges support of JP-FAPESP (2018/08845-3) and of CNPq-Universal (431083/2018-5). This work has also been supported by FAPESP (2019/26141-6). KB is supported by the Agence Nationale de la Recherche  (ANR-18-CE92-0020-01; ANR-19-CE30-0014-04). EBS acknowledges supports from FAPERJ under PVE E-26/101.468/2014, E-26/010.002990/2014 and E-261010.001246/201.

\appendix
\section{APPENDIX}
In this appendix we recall the relation between longitudinal velocity and elastic moduli, in particular for the case of an orthorhombic crystal structure. 

In a generic crystal, for a certain crystallographic direction, we can define three sound velocities (one longitudinal and two transverse), whose description is a function of the stress tensor \cite{kittel1966}. 
In a Cartesian representation, 9 stress components acting on the elementary cell can be generally considered: $X_x$, $X_y$, $X_z$, $Y_x$, $Y_y$, $Y_z$, $Z_x$, $Z_y$, $Z_z$. The capital letters point to the direction of the force, whereas the index indicates the direction which is perpendicular to the area where the force acts. As the total torque acting on the elementary cell has to be zero, the independent conditions are 6 as $Y_z= Z_y$, $Z_x= X_z$, $X_y= Y_x$.

In response to that external stress, the displacement of the deformation can be described by the strain $R(r)$:

\begin{equation}
R(r)=u(\textbf{r})\hat{x}+v(\textbf{r})\hat{y}+w(\textbf{r})\hat{z}
\end{equation}
 where $u(\textbf{r})$, $v(\textbf{r})$ and $w(\textbf{r})$ represent the displacements along the $x$, $y$ and $z$ axes, respectively. We define the strain components by the following relations: 
\begin{equation}
    e_{xx} \equiv \frac{\partial u}{\partial x} \hspace{0.4cm}
e_{yy} \equiv \frac{\partial v}{\partial y} \hspace{0.4cm}
e_{zz} \equiv \frac{\partial w}{\partial z}
\end{equation}
\begin{equation}
  e_{xy} \equiv \frac{\partial u}{\partial y}+\frac{\partial v}{\partial x}\hspace{0.2cm}
e_{yz} \equiv \frac{\partial v}{\partial z}+\frac{\partial w}{\partial y} \hspace{0.2cm}
e_{xy} \equiv \frac{\partial u}{\partial z}+\frac{\partial w}{\partial x}
\end{equation}

For small deformation, we can apply the Hooke's law and relate, at first approximation, the force (stress) to the strain through the elastic moduli $C_{\alpha \beta}$. For instance, for the $X_x$ stress component:
\begin{equation}
\begin{aligned}
X_x=C_{11}e_{xx}+C_{12}e_{yy}+C_{13}e_{zz}+C_{14}e_{yz}+C_{15}e_{zx}+C_{16}e_{xy}\\
\end{aligned}
\label{moduli}
\end{equation}

The 36 $C_{\alpha \beta}$ elastic constants, obtained when eq. \ref{moduli} is written for all directions, reduce to 21, as it can be shown that $C_{\alpha \beta}=C_{\beta \alpha}$. 

When the crystal symmetry is taken into account, the number of non-zero elastic constants can be reduced to 3 for a cubic system, 5 for an hexagonal system, and 9 for an orthorhombic. The latter is the case of interest in this work. 

The sound velocities of SnSe can be computed through the 9 non-zero elastic moduli of the stiffness tensor (see Table \ref{tab:matrixSnSe}), which were measured by ultrasound spectroscopy in single crystalline SnSe \cite{karunarathne2018}.

\begin{table}[h]
    \centering
    \begin{tabular}{ c | c  c  c  c c  c} 
   & $e_{xx}$ & $e_{yy}$ & $e_{zz}$ & $e_{yz}$ & $e_{zx}$ & $e_{xy}$\\ [0.5ex] 
   \hline
 $X_x$ & $C_{11}$ & $C_{12}$  & $C_{13}$  & 0 & 0 & 0\\ 
$Y_y$ & $C_{21}$ & $C_{22}$  & $C_{23}$  & 0 & 0 & 0\\ 
$Z_z$ & $C_{31}$ & $C_{32}$  & $C_{33}$  & 0 & 0 & 0\\ 
$Y_z$ & 0 & 0  & 0 & $C_{44}$ & 0 & 0\\ 
$Z_x$ & 0 & 0  & 0 & 0 & $C_{55}$ & 0\\ 
$X_y$& 0 & 0  & 0  & 0 & 0 & $C_{66}$
 \end{tabular}
    \caption{\textbf{Elastic constants:} The non-zero constant for an orthorhombic system. Numerical values are obtained from Ref. \cite{karunarathne2018} and expressed in GPa: $C_{11}=41.8$, $C_{22}=59.7$, $C_{33}=32.4$, $C_{44}=13.2$, $C_{55}=24.5$, $C_{66}=20.5$, $C_{12}=3.15$, $C_{13}=10.7$, $C_{23}=26.8$. } 
    \label{tab:matrixSnSe}
\end{table}

Then, we can relate the resultant force with the displacement, writing the motion equation. For instance, let us consider a travelling wave along [100], which corresponds to what was dubbed $x$-direction.
\begin{equation}
u(\textbf{r})=u_0exp[i(kx-\omega t)]
\end{equation} 
The motion equation is:
\begin{equation}
\rho \frac{\partial^2  u}{\partial t^2}= \frac{\partial X_x}{\partial x}+ \frac{\partial X_y}{\partial y}+ \frac{\partial X_z}{\partial z} \label{eq-u}
\end{equation}
and a similar expression is found for $v(\textbf{r})$ and $w(\textbf{r})$.


Let us substitute $u(\textbf{r})$ in equation (\ref{eq-u}). 
\begin{equation}
\omega^2 \rho=C_{11}k^2
\end{equation}
that leads to the the longitudinal velocity:
\begin{equation}
v_{lx}=\frac{\omega}{k}=\sqrt{\frac{C_{11}}{\rho}}
\end{equation}

Considering instead a transverse displacement ($v$ or $w$), with the wave propagating again along the $x$ direction, $v=v_0exp[i(kx-\omega t)]$, we can obtain the transverse velocity. 




 
 
 



With an analogous approach, the second transverse velocity can be obtained. Again, applying the calculation to the other two directions and solving the motion equations, we can complete table \ref{tab:soundVformula}. Finally, we can compute sound velocities for SnSe using the available elastic constants (Table \ref{tab:soundSnSe}).  The longitudinal velocities found here are used in the main text for the $b$ and $c$ crystallographic directions.



\begin{table}[h!]
    \centering
    	\begin{tabular}{||c | c | c | c||} 
		\hline
		Wave & Longitudinal  & I-Transverse & II-Transverse \\ 
	   direction & velocity & velocity   & velocity \\ 
		\hline\hline
		x & $\sqrt{\frac{C_{11}}{\rho}}$ & $v_{xy}=\sqrt{\frac{C_{66}}{\rho}}$ & $v_{xz}=\sqrt{\frac{C_{55}}{\rho}}$ \\ 
		\hline
		y & $\sqrt{\frac{C_{22}}{\rho}}$ & $v_{yx}=\sqrt{\frac{C_{66}}{\rho}}$ & $v_{yz}=\sqrt{\frac{C_{44}}{\rho}}$ \\
		\hline
		z & $\sqrt{\frac{C_{33}}{\rho}}$ & $v_{zx}=\sqrt{\frac{C_{55}}{\rho}}$ & $v_{zy}=\sqrt{\frac{C_{44}}{\rho}}$ \\
		\hline
	\end{tabular}
    \caption{\textbf{Sound velocities:} Anisotropic sound velocities for an orthorhombic structure..}
    \label{tab:soundVformula}
\end{table}



\begin{table}[h]
    \centering
    \begin{tabular}{| c | c | c | c |} 
 \hline
 Axis & $v_l$ & $v_{t1}$ & $v_{t2}$ \\ [0.5ex] 
           & [$m/s$] & [$m/s$]  & [$m/s$] \\ [0.5ex] 
 \hline
 $a$ & 2632 & 1843 & 2015\\ 
 \hline
 $b$ & 3146 & 1843 & 1479 \\
 \hline
 $c$ & 2317 & 2015 & 1479\\
 \hline
    \end{tabular}
    \caption{\textbf{SnSe sound velocities:} Longitudinal and transverse sound velocities of SnSe calculated starting from the elastic constant elements, along the main crystallographic axes.}
    \label{tab:soundSnSe}
\end{table}

The average sound velocity, which is often employed, can be related to the elastic moduli. Nevertheless, it is noteworthy to observe that this implies an approximation that we recall here below. 

The average sound velocity can be determined from the following formula \cite{huntington1958_elasticConstants}:
\begin{equation}
    v_s=\left( \frac{1}{3} \sum_{i=1}^3 \int_V \frac{1}{v_i^3} \frac{d \Omega}{4 \pi} \right)^{-1/3}
\end{equation}
where the integral considers the three velocities (longitudinal and two transverse) for a large number of directions in the space. The expression simplifies in the frequently used equation:
\begin{equation}
    v_s=\left( \frac{1}{3} \left[ \frac{2}{v_t^3} + \frac{1}{v_l^3}\right] \right)
    \label{eq:meanvelocity}
\end{equation}
that strictly holds only when the material is isotropic, which means when transverse ($v_t$) and longitudinal sound velocities ($v_l$) are invariant with the direction. The latter condition applies for instance in polycrystalline materials or isotropic glasses. In 1963, it was shown that a good approximation of the average sound velocity can be ontained through the so called VRHG approximation (Voigt, Reuss, Hill and Gilvarry), which establishes a criterium to connect the elastic tensor determined for a single crystal  with the average shear and bulk moduli determined for the same compounds but in the polycristalline form \cite{anderson1963_simplifiedDebye}. There, average transverse ($\bar{v_t}$) and longitudinal ($\bar{v_l}$) velocities are defined according to: $\bar{v_t}=\sqrt{G/\rho}$ and $\bar{v_l}=\sqrt{B/\rho} $, 
where $B$ and $G$ are respectively the Hill's average bulk and shear moduli of a polycristal isotropic sample that are obtained as a function of the stress tensor's components of a single crystal of the same compound: $B=B(C_{ij})$ and $B=B(S_{ij})$ \cite{karunarathne2018, anderson1963_simplifiedDebye}. $C_{ij}$ are the elastic constants and $S_{ij}=C_{ij}^{-1}$ are the elastic compliance constants. The elements of this tensor can be experimentally determined and so the average velocity  $ \bar{v}_m=\left( \frac{1}{3} \left[ \frac{2}{\bar{v}_s^3} + \frac{1}{\bar{v}_l^3}\right] \right)$.
$\bar{v}_s$ was shown to approximate closely $v_s$ for a large class of materials. Applying this recipe, the average sound velocity for polycrystalline SnSe can be calculated from the elastic moduli: 1803~m/s as in \cite{karunarathne2018}. It is important to observe that in SnSe it exists a significant difference among the average sound velocities and the direction-dependent sound velocities.


%

\end{document}